\documentclass[aip,reprint]{revtex4-1}

\usepackage[pdftex]{graphics}
\usepackage{amssymb,amsmath}

\begin{document}

\title{Vibrational strong coupling in liquid water from cavity molecular dynamics}
\author{Annina Z. Lieberherr}
\affiliation{Department of Chemistry, University of Oxford, Physical and Theoretical Chemistry Laboratory, South Parks Road, Oxford, OX1 3QZ, UK}
\author{Seth T. E. Furniss}
\affiliation{Department of Chemistry, University of Oxford, Physical and Theoretical Chemistry Laboratory, South Parks Road, Oxford, OX1 3QZ, UK}
\author{Joseph E. Lawrence}
\affiliation{Laboratory of Physical Chemistry, ETH Z\"urich, 8093 Z\"urich, Switzerland}
\author{David E. Manolopoulos}
%\email{david.manolopoulos@chem.ox.ac.uk}
\affiliation{Department of Chemistry, University of Oxford, Physical and Theoretical Chemistry Laboratory, South Parks Road, Oxford, OX1 3QZ, UK}
\begin{abstract}
We assess the cavity molecular dynamics method for the calculation of vibrational polariton spectra, using liquid water as a specific example. We begin by disputing a recent suggestion that nuclear quantum effects may lead to a broadening of polariton bands, finding instead that they merely result in anharmonic red shifts in the polariton frequencies. We go on to show that our simulated cavity spectra can be reproduced to graphical accuracy with a harmonic model that uses just the cavity-free spectrum and the geometry of the cavity as input. We end by showing that this harmonic model can be combined with the experimental cavity-free spectrum to give results in good agreement with optical cavity measurements. Since the input to our harmonic model is equivalent to the input to the transfer matrix method of applied optics, we conclude that cavity molecular dynamics cannot provide any more insight into the effect of vibrational strong coupling on the absorption spectrum than this transfer matrix method, which is already widely used by experimentalists to corroborate their cavity results.
\end{abstract}

\maketitle

\section{Introduction}

There is considerable current interest in the effect of an optical cavity on a wide range of experimental observables, including linear and non-linear\cite{xian+18pnas,ribe+18jpcl} infrared spectra, the conductivity and photoconductivity of semiconductors,\cite{orgu+15nm,naga+20acsnano,krai+20prl} and liquid phase chemical reaction rates.\cite{thom+16acie,thom+19science,lath+19acie,verg+19acie,lath-geor21jpcl} However, the effects of the cavity on many of these observables are still poorly understood and intensely debated.\cite{impe+21jcp,wies-xion21jcp,mischa} Reports of cavity effects on reaction rates are especially surprising, because the optical modes of the cavity are only expected to be very weakly coupled to any individual chemical reaction in a liquid sample,\cite{yang-cao21jpcl} and because a more strongly coupled collective excitation of multiple reactions to their transition states would be expected to have a prohibitive activation energy.\cite{camp+23arxiv} But one thing at least is clear: because the optical modes of the cavity interact directly with the dipole moment of the system within it, the cavity has a pronounced and easily understood effect on the dipole absorption spectrum of the system as a whole.

Li {\em et al.}\cite{li+20pnas} have recently used classical molecular dynamics simulations to explore this effect for a cavity containing liquid water. Their results for the effect of an optical cavity on the dipole absorption spectrum of the q-TIP4P/F water model\cite{habe+09jcp} were as follows. In the vibrational strong coupling regime, they found that a single resonant cavity mode splits the broad intramolecular O--H stretching band into narrower upper and lower polariton bands, the frequencies and relative intensities of which can be explained using a simple two-state harmonic model in which a single bright (symmetric) linear combination of $N$ equal frequency O--H oscillators is coupled to the cavity.\cite{li+20pnas} They also found more complicated polariton spectra when multiple cavity modes were considered; they then saw additional resonant interactions with the librational and H--O--H bending bands of the liquid and isolated non-resonant cavity bands that were shifted from the underlying cavity frequencies.

In addition to these dipole absorption results, Li {\em et al.}\cite{li+20pnas}  also considered the effect of the cavity on other experimental observables. After showing that the cavity cannot alter any static equilibrium property of the system in the classical limit, they considered two further dynamical properties: the centre-of-mass velocity and first-order dipole axis orientational correlation functions of the individual water molecules in their simulations. They found no evidence of any cavity effect on the velocity autocorrelation function, and only very little effect of the cavity on the orientational correlation function. (The frequency spectrum of their orientational correlation function did show a tiny peak due to the cavity at the same frequency as a polariton peak in the dipole absorption spectrum,\cite{li+20pnas} but we suspect that even that would disappear in a simulation with more than the $216$ water molecules they used in their simulations.)

The upshot of Li {\em et al.}'s initial calculations was thus that, while the cavity does have a significant and easily understood effect on the collective dipole absorption spectrum of the liquid within it, it has no effect on any static equilibrium properties and an entirely negligible effect on single-molecule dynamical properties, at least in the classical nuclear motion limit.\cite{li+20pnas} To explore whether these conclusions were robust to the inclusion of nuclear quantum effects, they performed a follow-up study\cite{li+22jpcl} of the same water model using the thermostatted ring polymer molecular dynamics (TRPMD) method.\cite{ross+14jcp} They found that the cavity had no discernible effect on the static dielectric constant of the liquid in either their classical or their quantum simulations, and that the Rabi splitting of the O--H stretching polariton bands was the same in both calculations. However, they did observe one interesting new effect: the polariton bands were found to be approximately two times broader in TRPMD than in classical MD. What was not clear from their study, as they were careful to stress,\cite{li+22jpcl} was whether this was a genuine quantum mechanical effect, such as a faster quantum polariton relaxation rate than a classical one, or simply an artifact of the approximations inherent in TRPMD. The desire to get to the bottom of this question using a more reliable quantum simulation method is what motivated us to undertake the present study.

To explain what we mean by a more reliable quantum simulation method, it is useful to begin by pointing out that the development of imaginary time path integral methods for the calculation of vibrational spectra has had a long and checkered history. The two earliest methods, centroid molecular dynamics\cite{cao-voth94jcp} (CMD) and ring polymer molecular dynamics\cite{crai-mano04jcp} (RPMD), are now known to have serious issues in the high-frequency stretching region. CMD suffers from a curvature problem that causes stretching bands to broaden and red-shift with decreasing temperature,\cite{witt+09jcp,witt+10jcp} whereas RPMD suffers from a resonance problem in which the high-frequency bands are contaminated by spurious contributions from the ring polymer internal modes.\cite{shig-naka08cpl,habe+08jcp} The TRPMD method used by Li {\em et al.}\cite{li+22jpcl} eliminates the resonance problem by damping out the spurious resonances with an internal mode thermostat, but this has the unfortunate side effect of causing spectral bands to broaden with decreasing temperature.\cite{ross+14jcp} It is thus not clear whether the broadening of the liquid water polariton bands observed by Li {\em et al.}\cite{li+22jpcl} in their TRPMD simulations was due to the thermostat or to a genuine quantum mechanical phenomenon.

In more recent work, Althorpe and co-workers have developed a quasi-centroid molecular dynamics (QCMD) method that eliminates the curvature problem of CMD,\cite{tren+19jcp,bens+20fd} we have developed a fast implementation of QCMD that pre-computes the quasi-centroid potential of mean force from a short path integral molecular dynamics simulation,\cite{flet+21jcp} and Kapil and co-workers have suggested machine learning the centroid potential of mean force at an elevated temperature where the high frequency stretches are still in their quantum ground states but the curvature problem has not yet become an issue.\cite{musi+22jcp} All three of these newer methods have been shown to give similar results for the room temperature vibrational spectrum of liquid water, without any obvious artefacts, so one can have some confidence that they are physically reasonable.\cite{lieb+23jcp} Methods such as our fast QCMD and Kapil's elevated temperature CMD are also less expensive than previous path integral methods because they simply involve running classical trajectories on a pre-computed centroid (or quasi-centroid) potential of mean force. This makes them especially well suited to the calculation of vibrational polariton spectra for reasons we shall explain in Sec.~II.

\section{Cavity molecular dynamics}

\subsection{Classical molecular dynamics}

As has been discussed extensively in recent papers, the vibrational Hamiltonian for a system in an optical cavity has the general form\cite{li+20pnas,cohen-tannoudji97}
\begin{equation}
{H} = H_{\rm sys}+H_{\rm cav},
\end{equation}
where $H_{\rm sys}$ is the bare system Hamiltonian
\begin{equation}
H_{\rm sys} = \sum_{i} {{\bf p}_i^2\over 2m_i}+V(\{{\bf q}_i\}),
\end{equation}
and $H_{\rm cav}$ contains the optical field and light-matter coupling terms:
\begin{equation}
H_{\rm cav} = \sum_{cd} \left[{1\over 2}{p}_{cd}^2+{1\over 2}\omega_c^2\left({q}_{cd}+{\mu_{d}(\{{\bf q}_i\})\over \omega_c\sqrt{\epsilon_0V_{\rm cav}}}\right)^2\right].
\end{equation}
Here the double sum is over the modes of the cavity with frequencies $\omega_c$ and polarisation directions $d$, $\mu_{d}(\{{\bf q}_i\})$ is the component of the electric dipole moment of the system in the direction $d$, $\epsilon_0$ is the vacuum permittivity, and $V_{\rm cav}$ is the volume of the cavity. When the light is propagating along the laboratory $z$ axis the sum over $d$ includes $x$ and $y$. 

In the absence of an optical cavity, the dipole absorption spectrum of a liquid can be calculated from its dipole moment time-derivative autocorrelation function
\begin{equation}
I(\omega) = {\beta\over 6c\epsilon_0V_{\rm sys}}\int_{-\infty}^{\infty}  e^{-i\omega t}\sum_{d=x,y,z} \left<\dot{\mu}_d(0)\dot{\mu}_d(t)\right>\,{\rm d}t.
\end{equation}
Here $I(\omega)=n(\omega)\alpha(\omega)$, where $n(\omega)$ is the frequency-dependent refractive index and $\alpha(\omega)$ is the Beer-Lambert absorption coefficient of the liquid, $V_{\rm sys}$ is its volume, $c$ is the speed of light in a vacuum, and $\beta=1/k_{\rm B}T$. The corresponding expression when the liquid is inside a cavity is\cite{li+20pnas}
\begin{equation}
I(\omega) = {\beta\over 4c\epsilon_0V_{\rm sys}}\int_{-\infty}^{\infty}  e^{-i\omega t}\sum_{d=x,y} \left<\dot{\mu}_d(0)\dot{\mu}_d(t)\right>\,{\rm d}t,
\end{equation}
where the $z$ component of the autocorrelation function has been omitted so as to model an experiment in which the absorption is measured in the $z$ direction (the direction in which light propagates inside the cavity). 

The only remaining issue from the point of view of using classical molecular dynamics to simulate the cavity spectrum is how to specify the light-matter coupling strength in Eq.~(3). This is an adjustable parameter because the volume $V_{\rm cav}$ of the cavity is not necessarily the same as the volume $V_{\rm sys}$ of the liquid within it. For example, the liquid might be separated from the cavity mirrors by slabs of a low permittivity dielectric such as silica to eliminate edge effects and validate the long-wavelength approximation inherent in Eq.~(3).\cite{li+20pnas} It is however clear that the factor that multiplies $\mu_{d}(\{{\bf q}_i\})$ in Eq.~(3) must be proportional to $1/\sqrt{V_{\rm sys}}$, because the mean squared fluctuations of the dipole moment of a liquid are size extensive.\cite{size-extensive} So if $\mu_{d}(\{{\bf q}_i\})$ were not scaled by $1/\sqrt{V_{\rm sys}}$, the formula would not be size consistent, and its predictions would depend on the system size. (In particular, they would depend on the size of the system in the $x$ and $y$ directions between the two cavity mirrors.)

These considerations suggest using the dimensionless parameter $R=V_{\rm sys}/V_{\rm cav}$ -- the fraction of the cavity containing liquid -- to control the light-matter coupling strength in the calculation. With this choice, Eq.~(3) can be re-written as
\begin{equation}
H_{\rm cav} = \sum_{cd} \left[{1\over 2}{p}_{cd}^2+{1\over 2}\omega_c^2\left( {q}_{cd}+\sqrt{R}\,{\nu_{d}(\{{\bf q}_i\})\over \omega_c}\right)^2\right],
\end{equation}
where
\begin{equation}
\nu_d(\{{\bf q}_i\}) = {\mu_d(\{{\bf q}_i\})\over\sqrt{\epsilon_0V_{\rm sys}}}
\end{equation}
is a scaled (system size independent) dipole moment with the dimensions of a mass-scaled coordinate times a frequency ($M^{1/2}LT^{-1}$). Note that, with this prescription, is not necessary to simulate a real  Fabry-P\'erot microcavity containing on the order of 10$^{10}$ water molecules. One can instead simply keep $R$ constant and simulate systems of increasing size $V_{\rm sys}$ with periodic boundary conditions until the cavity spectrum has converged -- in the same way as Eq.~(4) (which is also size-consistent) is typically used to calculate a cavity-free dipole absorption spectrum. 

\subsection{Quasi-centroid molecular dynamics}

Quantum mechanical effects in the nuclear motion can be included in the above framework by replacing classical molecular dynamics with a QCMD or (elevated temperature\cite{musi+22jcp}) CMD simulation, which is especially easy in the present context for two key reasons. The first is that the light-matter coupling potential in Eq.~(3) is a purely harmonic function of the light coordinates $\{{q}_{cd}\}$, and the second is that the system coordinates $\{{\bf q}_i\}$ only enter the light-matter coupling via the system dipole moment function $\mu_{d}(\{{\bf q}_i\})$. When this dipole moment is a linear function of the system coordinates, as it is in the q-TIP4P/F water model,\cite{habe+09jcp} the forces on the centroids of the system coordinates arising from the light-matter coupling become decoupled from the internal modes of the ring polymer and can be treated classically: there is no quantum contribution to the mean force acting on the centroid of any system coordinate from the light-matter coupling. Furthermore, because the light-matter coupling potential is a harmonic function of the $\{{q}_{cd}\}$, and because these coordinates do not enter the Hamiltonian anywhere else, the mean force on the centroid of each light coordinate is also purely classical. The same is true in QCMD if we define the quasi-centroids of the light coordinates to coincide with their centroids, which is entirely natural because these are one-dimensional simple harmonic oscillators that are not expected to suffer from the curvature problem. 

It follows from these observations that a fast CMD or QCMD calculation of a vibrational polariton spectrum is extremely simple. All one has to do is pre-compute the (elevated temperature\cite{musi+22jcp}) centroid or (target temperature\cite{flet+21jcp}) quasi-centroid potential of mean force of the cavity-free system in a short path integral molecular dynamics simulation, and then perform a purely classical molecular dynamics simulation of the dipole absorption spectrum using the Hamiltonian in Eq.~(1) with this potential of mean force replacing the classical potential $V(\{{\bf q}_i\})$. This is how we have performed the present calculations, using a pre-computed\cite{lieb+23jcp} room temperature quasi-centroid potential of mean force for the q-TIP4P/F water model in all of our QCMD simulations.

\subsection{Nuclear quantum effects}

Fig.~1 compares the classical and QCMD cavity molecular dynamics approximations to the polariton spectrum of the q-TIP4P/F water model in the case where a single cavity mode (with two polarisation directions) is in resonance with the intramolecular O--H stretching band. We have also used RPMD\cite{crai+mano04jcp,habe+08jcp} and TRPMD\cite{ross+14jcp} to calculate the in-cavity and out-of-cavity spectra and have included these results in the figure for comparison. All four calculations (classical, RPMD, TRPMD, and QCMD) were preformed for a box of 216 water molecules with periodic boundary conditions at room temperature and the experimental density. The RPMD, TRPMD, and quasi-centroid potential of mean force\cite{lieb+23jcp} calculations all used 32 ring polymer beads.

We increased the light-matter coupling parameter $R=V_{\rm sys}/V_{\rm cav}$ until the Rabi splitting $\Omega_R$ of the polariton bands in our classical simulation was close to the 715 cm$^{-1}$ reported by Li {\em et al.}\cite{li+22jpcl} The resulting value of $R$ was found to be 2.25. This is clearly unphysical because it implies a larger volume of water in the cavity than the volume of the cavity itself. We shall consider more realistic values of $R$ below, but we have used this value in all four panels of Fig.~1 so as to make a direct connection with Li {\em et al.}'s discussion of the impact of nuclear quantum effects on the polariton bands.\cite{li+22jpcl}   

The classical results in Fig.~1 are very similar to those reported by Li {\em et al.}\cite{li+22jpcl} In both our calculation and theirs, the resonant cavity is found to split the O--H stretching band asymmetrically into upper and lower polariton bands, with the upper polariton carrying the majority of the dipole intensity. Our (their) computed Rabi splitting is 720 (715) cm$^{-1}$, and the full widths at half maximum of our (their) upper and lower polariton bands are 18 (24) and 59 (60) cm$^{-1}$, respectively. The relative narrowness of the polariton bands arises because the coupling to the cavity eliminates the inhomogeneous broadening that is present in the cavity-free spectrum.\cite{houd+96pra} The asymmetry of the polariton bands can be explained using a simple harmonic oscillator model in which a single bright (symmetric) linear combination of degenerate O--H stretches is resonant with a single optical cavity mode.\cite{li+20pnas}

The TRPMD results in Fig.~1 are also very similar to those reported by Li {\em et al.}\cite{li+22jpcl} The TRPMD bands are red shifted and broadened relative to their classical counterparts in both the in-cavity and the out-of-cavity spectra. In our (their) TRPMD simulations, the Rabi splitting is 727 (720) cm$^{-1}$, and the full widths at half maximum of our (their) upper and lower polariton bands are 78 (65) and 134 (129) cm$^{-1}$, respectively. As we have already mentioned in Sec.~I, it is not clear whether the broadening of the polariton bands seen in TRPMD is a genuine quantum mechanical effect or an artefact of the internal mode thermostat that is used in this method to mitigate the RPMD resonance problem. 

\begin{figure}[t]
\centering
\resizebox{0.9\columnwidth}{!} {\includegraphics{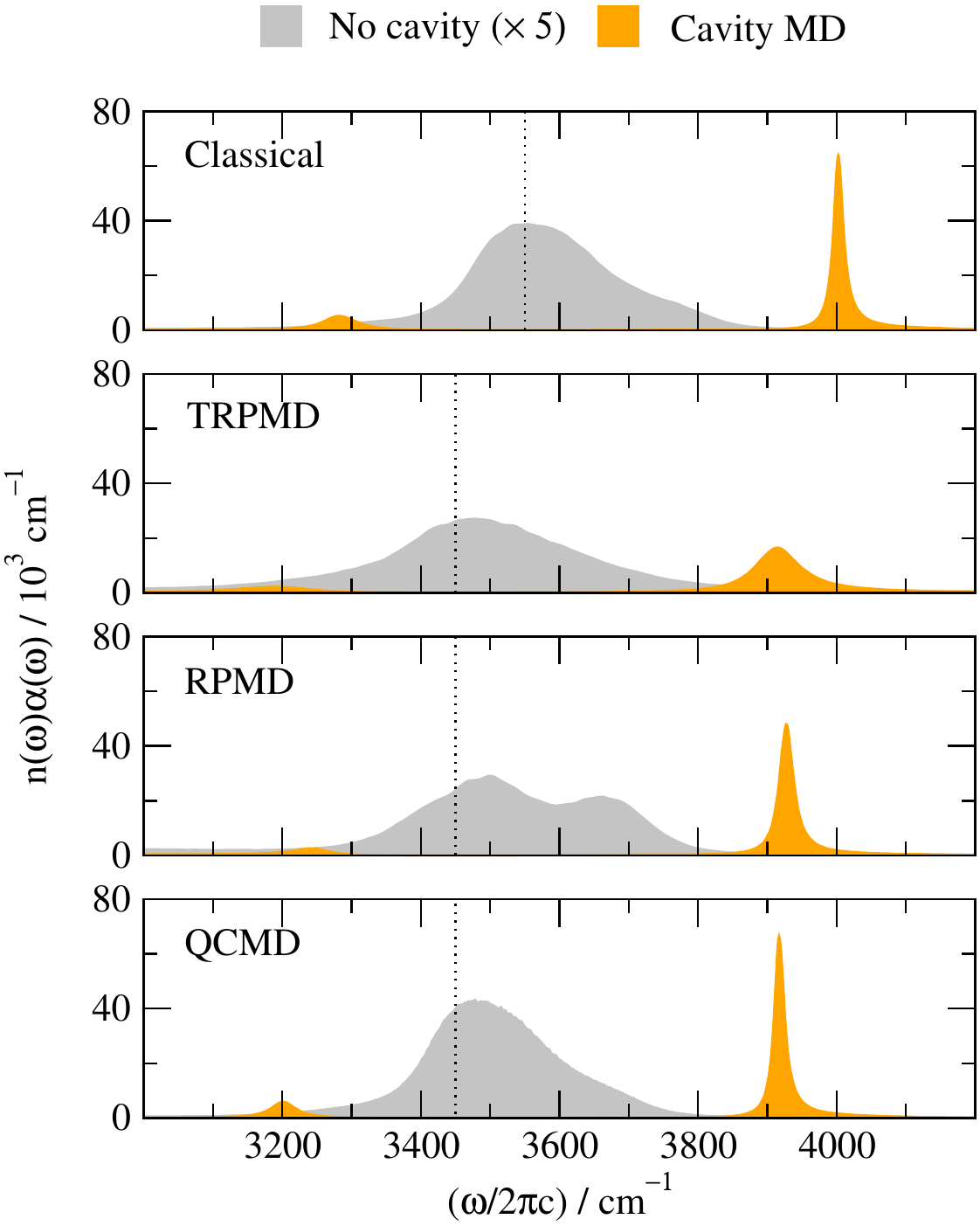}}
\caption{The effect of a single resonant cavity mode on the O--H stretching band of the q-TIP4P/F water model, as calculated using classical molecular dynamics and three different imaginary time path integral-based molecular dynamics methods for $R=V_{\rm sys}/V_{\rm cav}=2.25$. The cavity frequencies used in the simulations are indicated by the dotted vertical lines, and the intensities of the cavity-free spectra have been multiplied by a factor of 5 to make them easier to see. Before this factor of 5 is introduced the integrated intensities of the cavity-free and cavity MD spectra are the same within each panel.}
 \end{figure}

The RPMD results in Fig.~1 shed some light on this question. In the absence of an optical cavity, the RPMD spectrum is contaminated by spurious resonance interactions with the internal modes of the ring polymer.\cite{shig-naka08cpl,habe+08jcp} However, it seems that when the cavity is present this problem is fortuitously avoided, at least in the situation we are considering here. While the lower polariton band may still be affected by the resonance problem, the upper polariton band in the RPMD panel in Fig.~1 has a physically reasonable line shape that does not show any evidence of spurious resonance interactions. Note also that the line shape of this band is significantly narrower than in the TRPMD calculation. This suggests that the polaritons obtained by Li {\em et al.}\cite{li+22jpcl} were indeed broadened by the internal mode thermostat used in their TRPMD calculations.

Further evidence for this is provided by the final panel in Fig.~1, which shows the results of our present QCMD calculations. Here the line shapes of the in-cavity and out-of-cavity spectra are basically the same as they are in the classical simulation. The only quantum mechanical effect is a slight anharmonic red shift in the frequencies of the various bands (both cavity-free and polaritonic).  Since the cavity-free QCMD spectrum in Fig.~1 is free from any obvious artefacts, and since a consensus is now emerging in the literature that this is the correct O--H stretching spectrum of the room temperature q-TIP4P/F water model,\cite{bens+20fd,musi+22jcp,lieb+23jcp} we feel it is safe to conclude that nuclear quantum effects do not have any impact on the widths of the polariton bands. The only nuclear quantum effect that is seen in the QCMD results in Fig.~1 is a red shift of around 100 wavenumbers, in both the in-cavity and the out-of-cavity spectra. 

\subsection{More realistic simulations}

The Rabi splittings in Fig.~1 are unrealistically large because $R=V_{\rm sys}/V_{\rm cav}>1$ is not physically realisable. Figs.~2 and~3 show the results of cavity QCMD simulations of the q-TIP4P/F water model with $R=1.0$ and $R=0.5$, respectively. $R=1.0$ corresponds to the situation in which the cavity is filled with water and it therefore gives the largest Rabi splittings that the q-TIP4P/F model could feasibly produce. $R=0.5$ corresponds to the situation in which the volume of water in the cavity is half the volume of the cavity. This could be achieved (for example) by separating the water layer from the cavity mirrors by slabs of a low dielectric material such as silica.\cite{li+20pnas} 

The results shown in Figs.~2 and 3 are for four different scenarios: a single cavity mode resonant with the O--H stretch, a single cavity mode resonant with the H--O--H bend, a single cavity mode resonant with the librational band, and an example in which there are multiple cavity modes. 

\begin{figure}[t]
\centering
\resizebox{0.9\columnwidth}{!} {\includegraphics{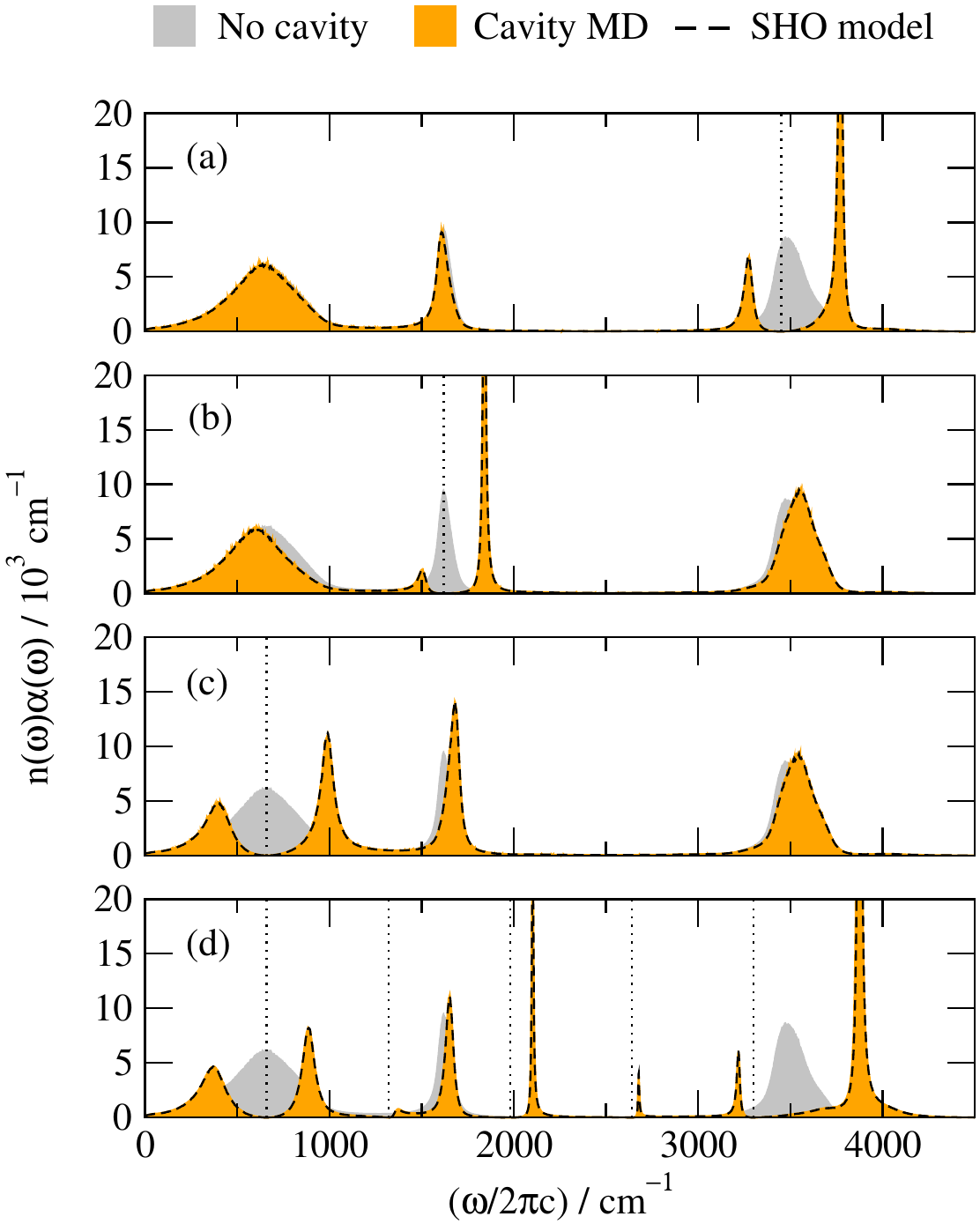}}
\caption{The effect of various optical cavities on the QCMD dipole absorption spectrum of the q-TIP4P/F water model. The results obtained from cavity molecular dynamics are compared with those obtained using the simple harmonic oscillator model described in Sec.~III. The cavity frequencies are indicated as dotted vertical lines in each panel: (a) a single resonant interaction with the O--H stretch; (b) a single resonant interaction with the H--O--H bend; (c) a single resonant interaction with the librational band; (d) an example with multiple cavity modes. $R=V_{\rm sys}/V_{\rm cav}=1.0$ in all four panels.}
 \end{figure}
 
\begin{figure}[t]
\centering
\resizebox{0.9\columnwidth}{!} {\includegraphics{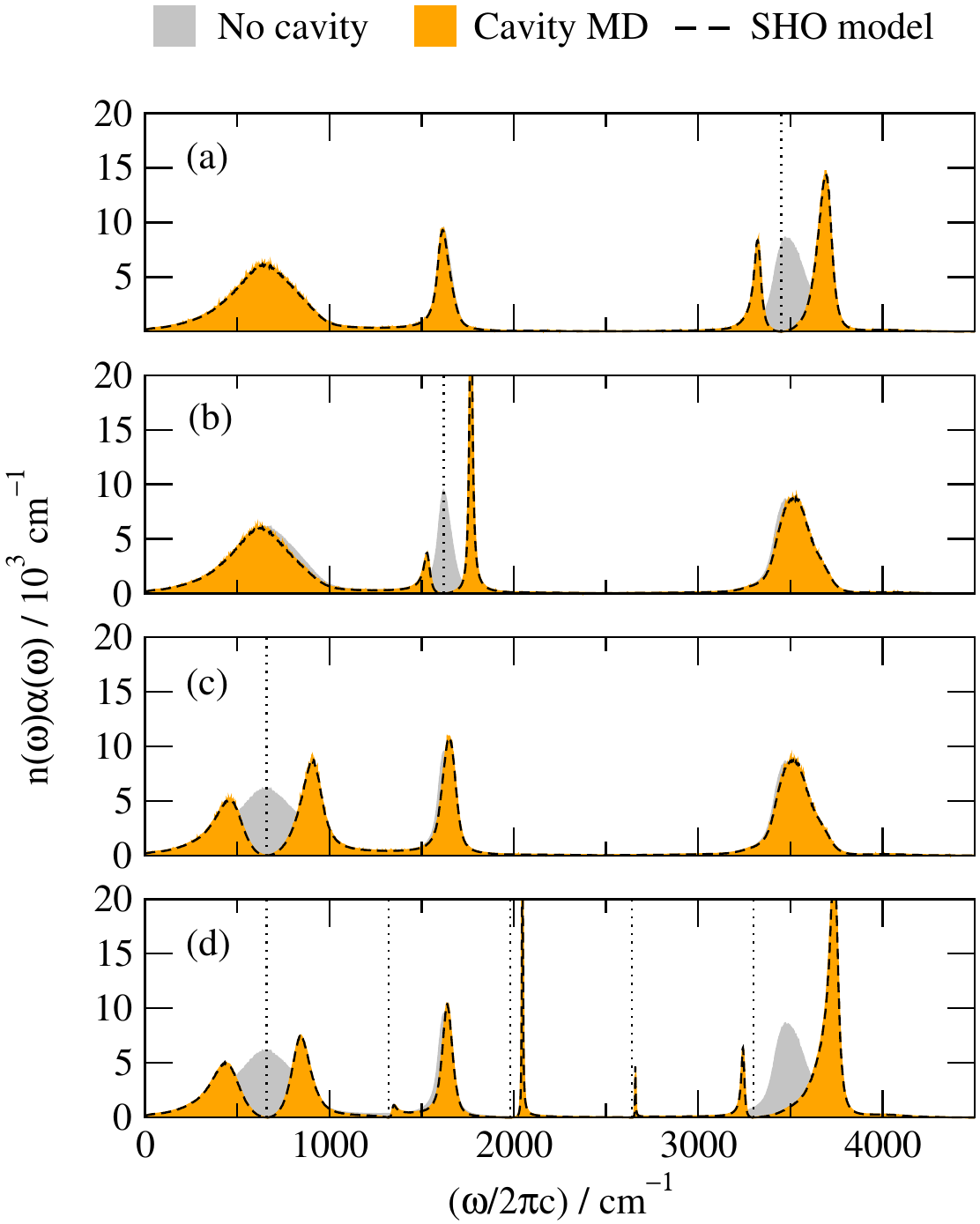}}
\caption{As in Fig.~3, but for $R=V_{\rm sys}/V_{\rm cav}=0.5$.} 
\end{figure}

Consider first the results in Fig.~2. Here the Rabi splittings are as large as they can possibly be, and they are manifestly in the vibrational strong coupling regime (i.e., larger than the widths of the cavity-free bands). The results for a single cavity mode resonant with the O--H stretch in panel~(a) can be compared directly with the QCMD results in panel (d) of Fig.~1. The Rabi splitting is smaller in Fig.~2 because the light-matter coupling strength is smaller, and the expanded frequency scale reveals that the resonance of the cavity mode with the O--H stretch has very little effect on the bending and librational bands (there is a tiny red shift of the bending band as a result of its repulsion by the lower stretching polariton, but this is barely noticeable in the figure). 

Similar effects are observed when the resonant cavity mode interacts with the bending and librational bands of the liquid as shown in panels (b) and (c). The band in resonance with the cavity is split asymmetrically into upper and lower polaritons, and the remaining bands in the spectrum are shifted slightly by the repulsion from these polaritons. The multiple cavity mode results in panel (d) are more complicated. Here there are a mixture of on-resonance interactions, near-resonance interactions, water band shifts, and isolated polariton bands that are shifted away from the underlying cavity frequencies. All of this is consistent with the classical results for a multi-mode cavity reported in the supplementary information of Li {\em et al.}'s initial paper.\cite{li+20pnas} The only difference is that the present simulations were performed with QCMD and therefore include quantum mechanical (anharmonic) shifts in the frequencies of the various bands.

Finally, consider Fig.~3, in which $R$ has been reduced by a factor of 2. The Rabi splittings are now smaller by a factor of approximately $\sqrt{2}$ because the light-matter coupling strength in Eq.~(6) is proportional to $\sqrt{R}$, but they are all still either in or approaching the vibrational strong coupling regime. Everything else in Fig.~3 is also as one would expect it to be for a smaller light-matter coupling strength, including the smaller shifts that are seen in the frequencies of the non-resonant bands. All of this makes good sense. But what is more interesting is that {\em all} of the effects of the various cavities on the spectra in Figs~2 and~3 can be captured using a simple harmonic oscillator model that eliminates the need to perform a cavity molecular dynamics simulation in the first place, as we shall describe next.

 \section{A simple harmonic oscillator model}
 
\subsection{Theory}
 
Since the $x$, $y$ and $z$ directions are all equivalent in an isotropic liquid, and the $x$ and $y$ directions remain equivalent when the liquid is in a cavity with mirrors reflecting light along the $z$ direction, we can simplify both the out-of-cavity and in-cavity expressions for the absorption spectrum in Eqs.~(4) and~(5) to
\begin{equation}
I(\omega) = {\beta\over 2c\epsilon_0V_{\rm sys}}\int_{-\infty}^{\infty}  e^{-i\omega t}\left<\dot{\mu}_x(0)\dot{\mu}_x(t)\right>\,{\rm d}t.
\end{equation}
This equation continues to hold in quantum mechanics when $\left<\dot{\mu}_x(0)\dot{\mu}_x(t)\right>$ is interpreted as the Kubo-transformed correlation function\cite{kubo57jpsj} that is approximated by methods like (T)RPMD and (Q)CMD. It can be written more compactly as
\begin{equation}
I(\omega) = {\beta\over 2c}\int_{-\infty}^{\infty} e^{-i\omega t}\left<\dot{\nu}_x(0)\dot{\nu}_x(t)\right>\,{\rm d}t ,
\end{equation}
where
$\nu_x = {\mu_x/\sqrt{\epsilon_0V_{\rm sys}}}$ is the $x$-component of the scaled (system size independent) dipole moment introduced in Eq.~(7).

Now suppose that the out-of-cavity spectrum $I(\omega)$ is available on a grid of ascending positive frequencies $\{\omega_i\}_{i=1}^N$. Then we can use the pre-limit delta functions
\begin{equation}
\delta_i(\omega) = \begin{cases} 1/\Delta\omega_i, & (\omega_{i-1}+\omega_i)/2 < \omega < (\omega_i+\omega_{i+1})/2\\ 0, & {\rm otherwise} \end{cases}
\end{equation}
where $\Delta\omega_i=(\omega_{i+1}-\omega_{i-1})/2$ with $\omega_0={\rm max}(0,2\omega_1-\omega_2)$ and $\omega_{N+1}=2\omega_N-\omega_{N-1}$, to construct a histogram approximation to the spectrum
\begin{equation}
I(\omega) \simeq  \sum_{i=1}^N \Delta\omega_iI(\omega_i)\left[\delta_i(\omega)+\delta_i(-\omega)\right],
\end{equation}
where we have exploited the fact that $I(\omega)$ is an even function of $\omega$. (This follows because both the classical and Kubo-transformed quantum mechanical correlation functions $\left<\dot{\mu}_x(0)\dot{\mu}_x(t)\right>$ are even functions of $t$.)

This histogram approximation can be used to map the cavity-free problem onto a simple harmonic oscillator (SHO) model as follows. The dipole time-derivative autocorrelation function of the SHO Hamiltonian
\begin{equation}
H_{\rm sys} = \sum_{i=1}^N {1\over 2}p_i^2+{1\over 2}\omega_i^2q_i^2
\end{equation}
with the (scaled) dipole moment function
\begin{equation}
\nu_x = \sum_{i=1}^N \nu_iq_i
\end{equation}
is
\begin{equation}
\left<\dot{\nu}_x(0)\dot{\nu}_x(t)\right> = {1\over\beta}\sum_{i=1}^N \nu_i^2\cos\omega_it,
\end{equation}
both classically and quantum mechanically (provided one again considers the Kubo-transformed correlation function in the quantum case). When this is substituted in place of $\left<\dot{\nu}_x(0)\dot{\nu}_x(t)\right>$ in Eq.~(9) it gives
\begin{equation}
I(\omega) = {\pi \over 2c}\sum_{i=1}^N \nu_i^2\left[\delta(\omega-\omega_i)+\delta(\omega+\omega_i)\right],
\end{equation}
the smoothed (histogram) version of which 
\begin{equation}
I(\omega) = {\pi \over 2c}\sum_{i=1}^N \nu_i^2\left[\delta_i(\omega)+\delta_i(-\omega)\right]
\end{equation}
will agree with Eq.~(11) if we set
\begin{equation}
\nu_i = \sqrt{2c\Delta\omega_iI(\omega_i)\over \pi},
\end{equation}
which has the dimensions of a frequency ($T^{-1}$).

Now consider what happens when we introduce an optical cavity. In view of Eq.~(6), the SHO Hamiltonian in Eq.~(12) will be augmented with a cavity Hamiltonian of the form (note that here we are considering a single polarisation direction $d=x$ so we can dispense with the sum over $d$),
\begin{equation}
H_{\rm cav}=\sum_{c=1}^{M} \left[{1\over 2}{p}_{c}^2+{1\over 2}\omega_c^2\left( {q}_{c}+\sqrt{R}\sum_{i=1}^N {\nu_iq_i\over\omega_c}\right)^2\right].
\end{equation}
The combined Hamiltonian $H=H_{\rm sys}+H_{\rm cav}$ is thus
\begin{equation}
H = {1\over 2}{\bf p}^T{\bf p}+{1\over 2}{\bf q}^T{\bf H}\,{\bf q}
\end{equation}
and the combined (scaled) dipole moment $\nu_x$ is
\begin{equation}
\nu_x = \boldsymbol{\nu}^T{\bf q},
\end{equation}
where ${\bf p} = (\{p_i\},\{p_c\})$, ${\bf q} = (\{q_i\},\{q_c\})$, $\boldsymbol{\nu} = (\{\nu_i\},\{0\})$, and the elements of the Hessian matrix ${\bf H}$ are $H_{ii'}=\omega_i^2\delta_{ii'}+MR\,\nu_i\nu_{i'}$, $H_{ic}=H_{ci}=\sqrt{R}\,\nu_i\omega_c$, and $H_{cc'}=\omega_c^2\delta_{cc'}$. 

Diagonalising the Hessian matrix with an orthogonal transformation matrix ${\bf C}$ 
\begin{equation}
{\bf C}^T{\bf H}\,{\bf C} = \tilde{\boldsymbol{\omega}}^2,
\end{equation}
and transforming the dipole moment vector $\boldsymbol{\nu}$ into the normal mode basis 
\begin{equation}
 {\bf C}^T\boldsymbol{\nu}=\tilde{\boldsymbol{\nu}},
\end{equation} 
we find that the in-cavity problem has the same structure as the out-of-cavity problem and so can be solved in the same way. Eq.~(14) becomes 
\begin{equation}
\left<\dot{\nu}_x(0)\dot{\nu}_x(t)\right> = {1\over\beta}\sum_{i=1}^{N+M} \tilde{\nu}_i^2\cos\tilde{\omega}_it,
\end{equation}
and the smoothed version of the in-cavity spectrum is therefore
\begin{equation}
\tilde{I}(\omega) = {\pi \over 2c}\sum_{i=1}^{N+M} \tilde{\nu}_i^2\left[\tilde{\delta}_i(\omega)+\tilde{\delta}_i(-\omega)\right],
\end{equation}
where $\tilde{\delta}_i(\omega)$ is defined by analogy with Eq.~(10):
\begin{equation}
\tilde{\delta}_i(\omega) = \begin{cases} 1/\Delta\tilde{\omega}_i, & (\tilde{\omega}_{i-1}+\tilde{\omega}_i)/2 < \omega < (\tilde{\omega}_i+\tilde{\omega}_{i+1})/2\\ 0, & {\rm otherwise} \end{cases}
\end{equation}
with $\Delta\tilde{\omega}_i=(\tilde{\omega}_{i+1}-\tilde{\omega}_{i-1})/2$, $\tilde{\omega}_0={\rm max}(0,2\tilde{\omega}_1-\tilde{\omega}_2)$, and $\tilde{\omega}_{N+1}=2\tilde{\omega}_N-\tilde{\omega}_{N-1}$. 

Within the SHO model, the effect of the cavity on the spectrum can thus be predicted using just the cavity-free spectrum and the geometry of the cavity as input. The cavity-free spectrum determines the $\omega_i$s in Eq.~(12) and the $\nu_i$s in Eq.~(17), and the geometry of the cavity determines both the cavity frequencies $\omega_c$ and the dimensionless parameter $R=V_{\rm sys}/V_{\rm cav}$ that controls the strength of the light-matter coupling term in Eq.~(18).
 
\begin{figure}[t]
\centering
\resizebox{0.9\columnwidth}{!} {\includegraphics{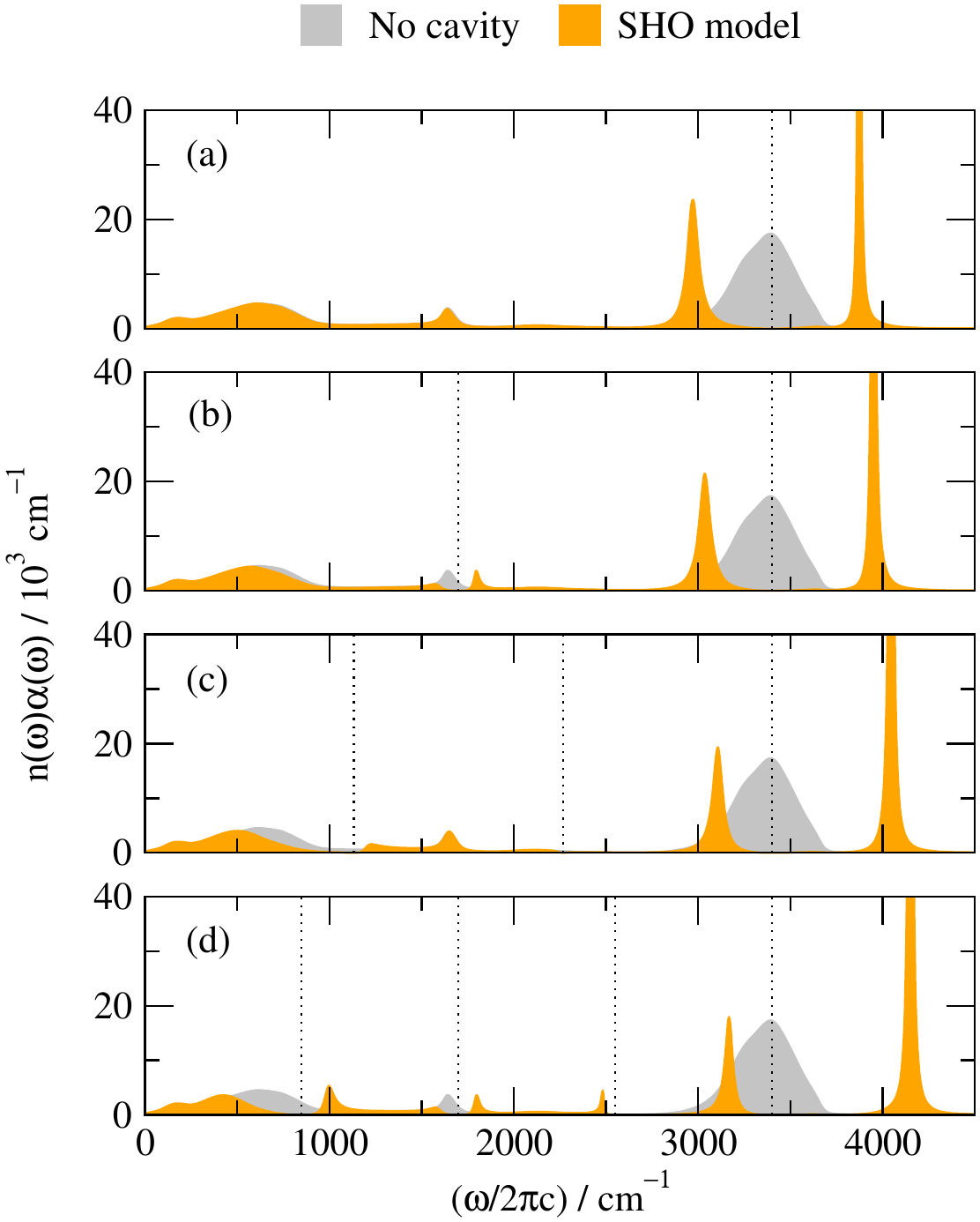}}
\caption{The effect of various optical cavities with frequencies up to $\omega_c/2\pi c=3400$ cm$^{-1}$ on the experimental dipole absorption spectrum of liquid water.\cite{bert-lan96as} The cavity modes with (a) $c=1$, (b) $c=2$, (c) $c=3$, and (d) $c=4$ are resonant with the O--H stretch. The cavity frequencies are indicated by the dotted vertical lines and the volume ratio is $R=V_{\rm sys}/V_{\rm cav}=1.0$.}
 \end{figure}
 
\subsection{Validation against cavity molecular dynamics}

We have used this SHO model to calculate the effects of various optical cavities on the dipole absorption spectrum of the q-TIP4P/F water model, using our cavity-free QCMD spectrum as input. In fact, we have already shown the results of some of these calculations in Figs~2 and~3 (as dashed black lines). These results reproduce those of our cavity QCMD simulations (the filled orange curves) almost perfectly: the only noticeable difference is that the curves from the SHO model are smoother than those from the simulations because they are free from statistical errors. Since this is true for a variety of different cavity frequencies and two different $V_{\rm sys}/V_{\rm cav}$ volume ratios, we would argue that the SHO model provides an entirely valid and considerably cheaper alternative to performing a cavity molecular dynamics simulation.
 
 \subsection{Application to real liquid water}
 
Another advantage of the SHO model over cavity molecular dynamics is that it can be used with {\em any} cavity-free spectrum as input, including the experimental cavity-free spectrum of the system of interest. This provides a way to eliminate all of the approximations that are made when performing molecular dynamics simulations, such as the use of an approximate interaction potential, the validity of the approximate treatment of nuclear quantum effects, and concerns about whether or not the simulation results are converged with respect to system size. 
 
The experimental (cavity-free) vibrational spectrum of liquid water is well known and has been available in digital form for many years.\cite{bert-lan96as}  We have therefore simply replaced our cavity-free QCMD spectrum of the q-TIP4P/F model with this experimental spectrum and repeated our SHO calculations. Selected results are shown in Fig.~4 for a cavity filled with water ($R=1.0$) when the cavity modes with $c=1$, 2, 3, and 4 are in resonance with the O--H stretching band.

The main difference between the experimental cavity-free spectrum and the q-TIP4P/F cavity-free spectrum is that the O--H stretching band is significantly more intense in reality than the q-TIP4P/F dipole moment function would suggest it to be.  Because of this, the light-matter coupling to the O-H stretch is stronger in reality than it is in the q-TIP4P/F model, the polariton spectra are more intense in the O--H stretching region, and the Rabi splittings of the O--H polaritons in Fig.~4 are larger than those in Fig.~2. 

One also sees from Fig.~4 that, on resonance, the Rabi splittings of the O--H polaritons are roughly the same no matter which cavity mode ($c=1$, 2, 3, or 4) is resonant with the O--H stretch, and that there is a slight shift of both polaritons to higher frequency with increasing $c$. The approximate independence of the Rabi splitting on the cavity mode is consistent with experimental measurements of water in optical cavities.\cite{impe+21jcp} The shift to higher frequencies with increasing $c$ is less obvious in experimental measurements and may be an artefact of us not having included higher frequency cavity modes in our simulations. There are also some subtler effects in the librational and bending regions in Fig.~4, but we suspect given their lower intensity that they would be more difficult to detect experimentally.

\begin{figure}[t]
\centering
\resizebox{0.8\columnwidth}{!} {\includegraphics{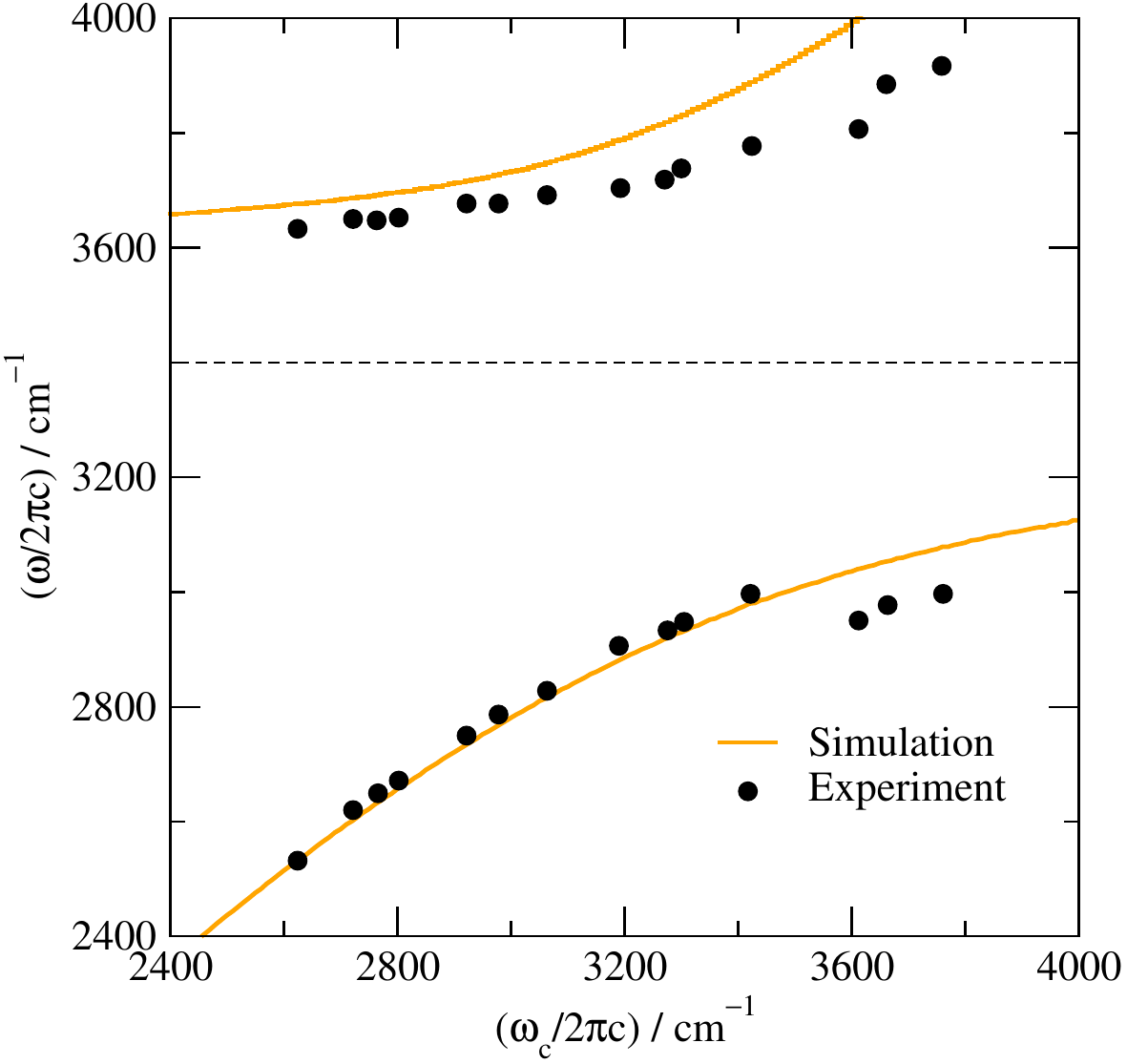}}
\caption{Comparison of simulated (SHO) and experimental (Ref.~\onlinecite{fuku+22jacs}) upper and lower polariton peak frequencies $\omega$ as a function of the resonant cavity frequency $\omega_c$, for the situation in which the cavity is almost completely filled with liquid water ($R=0.96$). The dashed horizontal line indicates the peak cavity-free O--H stretching frequency.}
 \end{figure}

\subsection{Comparison with a cavity experiment}

Vibrational strong coupling to the O--H stretching band of liquid water has been seen in a variety of experimental studies,\cite{verg+19acie,lath-geor21jpcl,impe+21jcp,fuku+22jacs} the results of which are qualitatively consistent with those in Fig.~4. To give just one quantitative example, we have used the SHO model with the experimental cavity-free spectrum as input to simulate the results of Fukushima {\em et al.},\cite{fuku+22jacs} who measured the upper and lower polariton peak positions as a function of the cavity frequency in the case where a single cavity mode with $c=1$ (corresponding to panel (a) in Fig.~4) was tuned through resonance with the O--H stretching band. 

The resulting comparison with experiment is shown in Fig.~5. We used $R=V_{\rm sys}/V_{\rm cav}=0.96$ in the simulations to allow for the fact that the water sample in the experiment was separated from the cavity mirrors by two narrow ($0.02$ $\mu$m) layers of SiO$_2$.\cite{fuku+22jacs} The cavity width was adjusted between 1.0 and 1.6 $\mu$m in the experiment so as to scan the cavity frequency, but we have not taken the implied adjustment of $R$ into account in our calculations. Nor have we taken into account the effect of the SiO$_2$ layers on the experimental cavity spectra, or the change in refractive index at the interfaces between these layers and the water. The quantitative agreement between theory and experiment in Fig.~5 is clearly reasonable nevertheless, especially for the lower polariton band. Since the upper polariton band is the narrower of the two (see Fig.~4), its peak position might perhaps be more sensitive to cavity loss processes, which we have also not taken into account in our calculations.

On the other hand, it is also possible that there is an issue with the experimental upper polariton results in Fig.~5. To investigate this, we have repeated the SHO model calculations for heavy water, again using the experimental cavity-free spectrum\cite{bert+89jpc} as input. The results are compared with the heavy water polariton spectra of Fukushima {\em et al.}\cite{fuku+22jacs} in Fig.~6. Here one sees that the agreement between the simulation and experiment is equally good for both of the polariton bands.

\begin{figure}[t]
\centering
\resizebox{0.8\columnwidth}{!} {\includegraphics{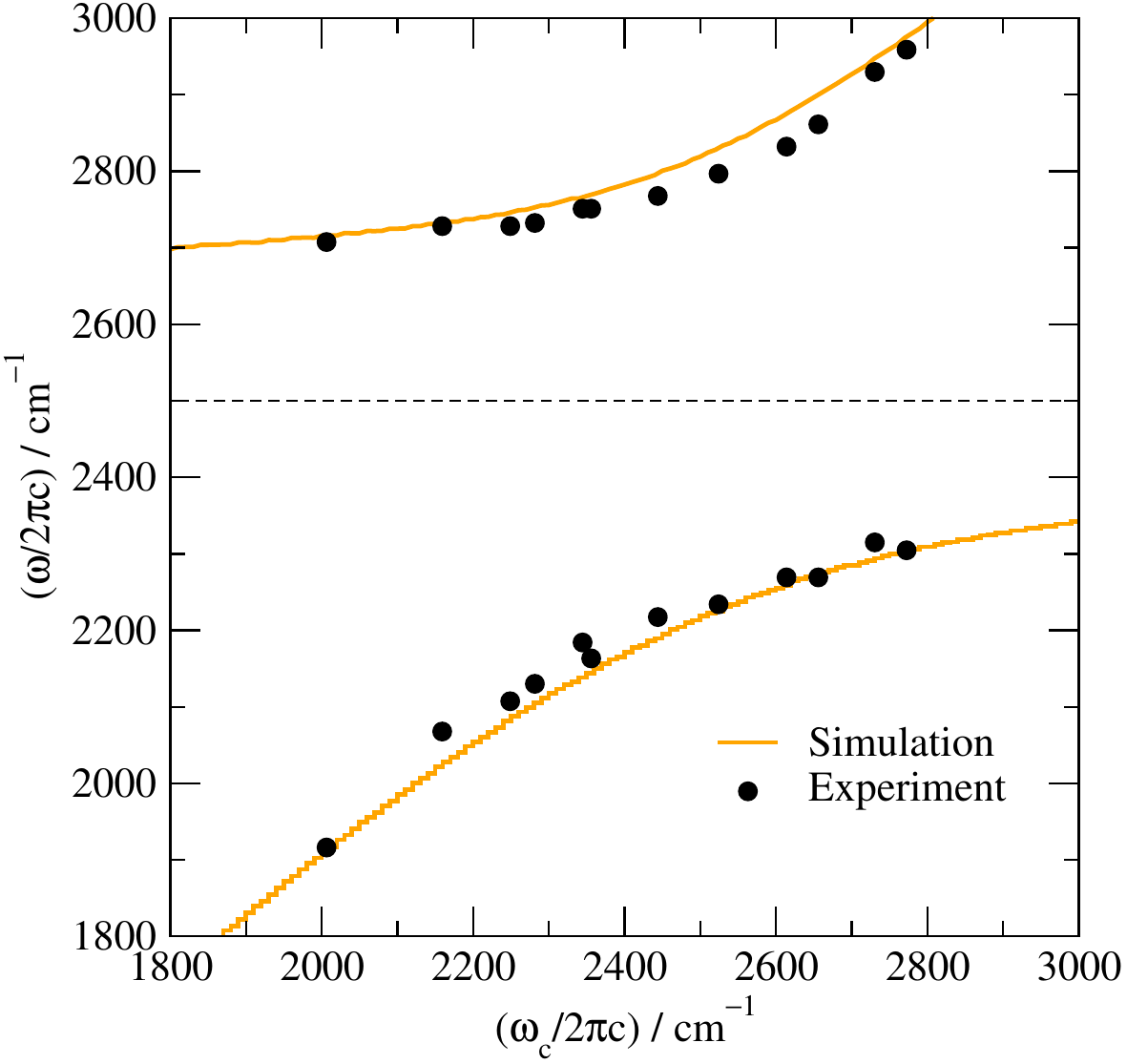}}
\caption{As in Fig.~5, but for heavy water. Here the experimental results are from the supplementary information of Ref.~\onlinecite{fuku+22jacs}, and the dashed horizontal line indicates the peak cavity-free O--D stretching frequency.}
 \end{figure}
 
\section{Concluding remarks}

In this paper, we have argued that:
\begin{itemize}
\item[(1)] There is no reliable evidence for any quantum mechanical broadening of the polariton bands in liquid water. Nuclear quantum effects do however result in anharmonic contributions to polariton band frequencies, as can be seen by comparing the first and last panels of Fig.~1.
\item[(2)] Cavity molecular dynamics spectra can be reproduced to graphical accuracy using the simple harmonic oscillator model introduced in Sec.~III, which only requires the cavity-free spectrum and the geometry of the cavity as input. (See Figs.~2 and 3.)
\item[(3)] Experimental measurements of vibrational strong coupling in liquid water are captured perfectly well by this SHO model when the experimental cavity-free spectrum is used as input. (See Fig.~5 and in particular Fig.~6.)
\end{itemize}
 
We feel that the second of these points is the most interesting, because it suggests that the only physical property of the material in the cavity that is needed to calculate the polariton spectrum is its (cavity-free) absorption spectrum. This is consistent with the use of the transfer matrix method from applied optics\cite{cent05ao,neme+20pra} to calculate the cavity spectrum, as is often done by experimental groups in this field to corroborate their results.\cite{thom+19science,lath+19acie,impe+21jcp,wies-xion21jcp,mischa}

The input to the transfer matrix method is the geometry of the cavity and the complex refractive index $N(\omega)=n(\omega)+ik(\omega)$ of each layer of material within it.\cite{cent05ao} When there is a single layer of material in the cavity, this is entirely equivalent to the spectral input $I(\omega)=n(\omega)\alpha(\omega)$ to our SHO model, because $n(\omega)$ and $k(\omega)$ are related by Kramers-Kronig relations and because the absorption coefficient is related to the extinction coefficient by $\alpha(\omega)=2\omega k(\omega)/c$.  When there are more layers of material in the cavity, the transfer matrix method accounts for the optical transmission through each successive layer and the change in refractive index on passing from one layer to the next.\cite{cent05ao} Its treatment of the cavity mirrors can also be adapted to allow for (linear) cavity loss.\cite{neme+20pra} None of these features is so easy to include in either our SHO model or a cavity molecular dynamics simulation. Moreover neither of these is expected to add anything useful to a transfer matrix calculation even for a single layer, because they do not incorporate any additional information (given point (2) above). 

Insofar as dipole absorption spectroscopy is concerned, we would therefore conclude that there is nothing a cavity molecular dynamics simulation has to offer over a standard transfer matrix calculation, which provides a simpler and more generally applicable way to corroborate experimental cavity results.\cite{thom+19science,lath+19acie,impe+21jcp,wies-xion21jcp,mischa} We would also repeat that Li {\em et al.}\cite{li+20pnas,li+22jpcl} found no significant evidence of any cavity effect on any single-molecule (non-collective) thermal equilibrium property of water in their simulations, as is to be expected from the extremely weak coupling ($\propto 1/\sqrt{N}$) of the cavity modes to individual molecules. There may however be significant cavity effects in non-equilibrium situations that cannot be described so well by linear response theory, such as photo-initiated processes that start out far from equilibrium as a result of vibration-polariton pumping.\cite{li+22nc} So there is certainly still a potentially useful role for cavity MD in shedding light on these non-equilibrium situations.

\section*{Supplementary Material}

\noindent
The supplementary material contains a listing of the computer program we used to generate the SHO results in Figs.~2-6, along with example input and output files.

\section*{Acknowledgements}

\noindent
We are grateful to the authors of Ref.~\onlinecite{li+22nc} for a helpful comment on the original draft of this manuscript. Annina Lieberherr is supported by a Berrow Foundation Lord Florey Scholarship. 

\section*{Author declarations}

\subsection*{Conflict of interest}

\noindent
The authors have no conflicts to disclose.

%\subsection*{Author contributions}

%\noindent
%{\bf Annina Lieberherr} instigated this project. She and {\bf Seth Furniss} performed the calculations. {\bf Joseph Lawrence} provided the potential of mean force for the QMCD simulations. {\bf David Manolopoulos} supervised the project and wrote the paper.

\section*{Data Availability}

\noindent
The data that support the findings of this study are available within the article and its supplementary material.


\begin{thebibliography}{99}
\bibitem{size-extensive}
Note that if this were not the case it would not be possible to calculate the dielectric constant of the liquid as
$\epsilon_r = 1+(4\pi\beta/3 V_{\rm sys})\Bigl[\left<\boldsymbol{\mu}\cdot\boldsymbol{\mu}\right>-\left<\boldsymbol{\mu}\right>\cdot\left<\boldsymbol{\mu}\right>\Bigr].$
\end{thebibliography}
\end{document}